\documentclass[reprint, amsmath,amssymb,
 aps,superscriptaddress]{revtex4-2}
\usepackage{graphicx} 
\usepackage{bm}
\usepackage{natbib}
\usepackage{verbatim}
\usepackage{epstopdf}
\usepackage{hyperref}
\usepackage{xcolor}
\usepackage{amsmath}
\usepackage[section]{placeins}
\usepackage{amsmath}
\usepackage{amssymb}
\usepackage{graphicx}

\begin{document}

	\title{A current source with metrological precision made on a 300mm silicon MOS process}
	\author{Nathan Johnson}
	\email{nathan.johnson@ucl.ac.uk}
	\affiliation{London Centre for Nanotechnology, University College London, 17-19 Gordon Street, London, WC1H 0AH, United Kingdom}
	
    \author{Stefan Kubicek}
	\affiliation{IMEC, Kapeldreef 75, 3001 Leuven, Belgium}

    \author{Julien Jussot}
	\affiliation{IMEC, Kapeldreef 75, 3001 Leuven, Belgium}

    \author{Yann Canvel}
	\affiliation{IMEC, Kapeldreef 75, 3001 Leuven, Belgium}

    \author{\\Kristiaan De Greve}
	\affiliation{IMEC, Kapeldreef 75, 3001 Leuven, Belgium}

	\author{M. Fernando Gonzalez-Zalba}
    \affiliation{Quantum Motion, 9 Sterling Way, London, N7 9HJ, United Kingdom}
	
    \author{Ross C. C. Leon}
    \affiliation{Quantum Motion, 9 Sterling Way, London, N7 9HJ, United Kingdom}
	
	\author{John J. L. Morton}
	\affiliation{London Centre for Nanotechnology, University College London, 17-19 Gordon Street, London, WC1H 0AH, United Kingdom}
	\affiliation{Quantum Motion, 9 Sterling Way, London, N7 9HJ, United Kingdom}

	\date{\today}

\begin{abstract}
	Although the measurement of current is now defined with respect to the electronic charge, producing a current standard based on a single-electron source remains challenging. The error rate of a source must be below 0.01~ppm, and many such sources must be operated in parallel to provide practically useful values of current in the nanoampere range.
    Achieving a single electron source using an industrial grade 300~mm wafer silicon metal oxide semiconductor (MOS) process could offer a powerful route for scaling, combined with the ability for integration with control and measurement electronics. 
	Here, we present measurements of such a single-electron source
    indicating an error rate of 0.008~ppm, below the error threshold to satisfy the SI Ampere, and one of the lowest error rates reported,
 	implemented using a gate-defined quantum dot device fabricated on an industry-grade silicon MOS process. 
	Further evidence supporting the accuracy of the device is obtained by comparing the device performance to established models of quantum tunnelling, which reveal the mechanism of operation of our source at the single particle level.
	The low error rate observed in this device motivates the development of scaled arrays of parallel sources utilising Si MOS devices to realise a new generation of metrologically accurate current standards.
\end{abstract}

	\maketitle

The 2019 redefinition of the International System of units (SI) determined a number of physical constants to have precise values ~\cite{BIPM, Stock}. 
The unit of electrical current, the Ampere, is now defined
as a number of elementary charges $e$ moving in a second $s$, and thus replaces the cumbersome definition that referenced the Ampere to the magnetic force between two currents passing parallel wires ~\cite{Giblin, PekolaReview}.
Semiconductors provide a natural platform to realise this new current standard, where there has been significant progress in the manipulation of single-electron currents~\cite{Philips, Fletcher3, Ed, Yamahata5, Schreiber}.
However, realising this new Ampere at the required metrological precision, with an error below 0.01~ppm (or $10^{-8}$) in the charges transferred individually per unit time, has proven challenging~\cite{PekolaReview, Kaestner3}.
Whilst there have been attempts to utilise the bulk flow of charge to realise a current source \cite{Rodenbach, deGraaf,Poirier}, quantum dot (QD) based sources have been promising both in precision and in applications such as forming nano-electronic circuits, or in closing the quantum metrological triangle \cite{Keller3, Scherer}. 

Quantum dot single-electron sources have been demonstrated in several materials including GaAs, graphene and silicon ~\cite{Blumenthal, Connolly, Yamahata, Giblin,GiblinReview}. In particular devices based on GaAs/AlGaAs heterostructures and Si nanowires have shown error rates close to the required 0.01~ppm level at frequencies approaching 1~GHz, generating currents at the $\sim 100$~pA level \cite{GiblinReview}.
To be of practical benefit, QD current sources aiming to realize the current standard should produce (i) a current on-demand at the nanoampere level ~\cite{PekolaReview, Stock}, which may require the parallelization of multiple QD sources, (ii) be embeddable with other electronic elements such as microwave generators and transimpedance
amplifiers that would complement the current standard and (iii) deployable without using a calibration chain. 
Such requirements have motivated work on silicon-based QD sources \cite{Jehl, Yamahata4, Rossi} over GaAs that lacks the maturity of silicon as a large scale semiconductor manufacturing platform. 
Further, GaAs-based devices typically require high magnetic fields (10~-~14~T).

Recent results report the co-integration of single-electron sources with on-chip signal generation and addressing electronics \cite{Dash} and the demonstration of parallel operation of up to four Si nanowire devices~\cite{Yamahata7}.
These results demonstrate the potential of SiMOS QDs as a platform for parallelized current sources, and how moving to industrial grade fabrication processes could enable integrated metrologically-precise on-chip current generation.
However, in both cases above the precision falls several orders of magnitude short of the required $10^{-8}$ error.

Here, we demonstrate the significant potential of current sources based on industrial-grade 300~mm wafer silicon CMOS, showing evidence of metrological precision at zero applied magnetic field. 
The device benefits from significant tunability, allowing for potential scalability by overcoming recently reported limitations  \cite{Yamahata7}, enabling parallelisation of devices and simplification, for example by using shared voltage schemes, in addition to the advantages of industrial CMOS processing.
Our study encourages further work to leverage the large-scale integration capabilities of silicon to (i) produce high current sources by parallelization of homogenous devices and (ii) integration with additional support electronics.
Further, we present a detailed analysis method and understanding of the single-electron transfer process in our device.

\begin{figure*}[th]
	\includegraphics[scale=0.3]{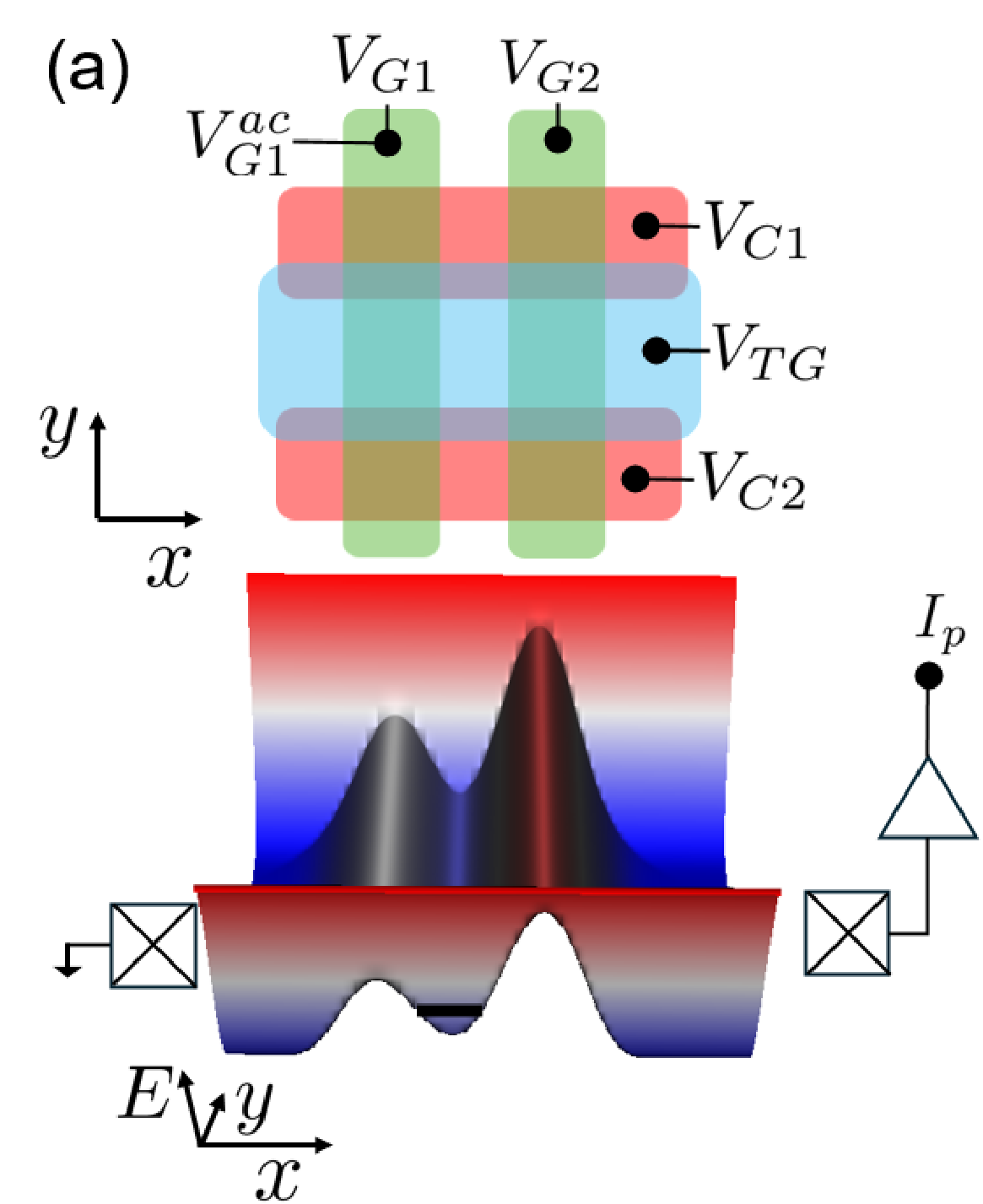}
	\includegraphics[scale=0.3]{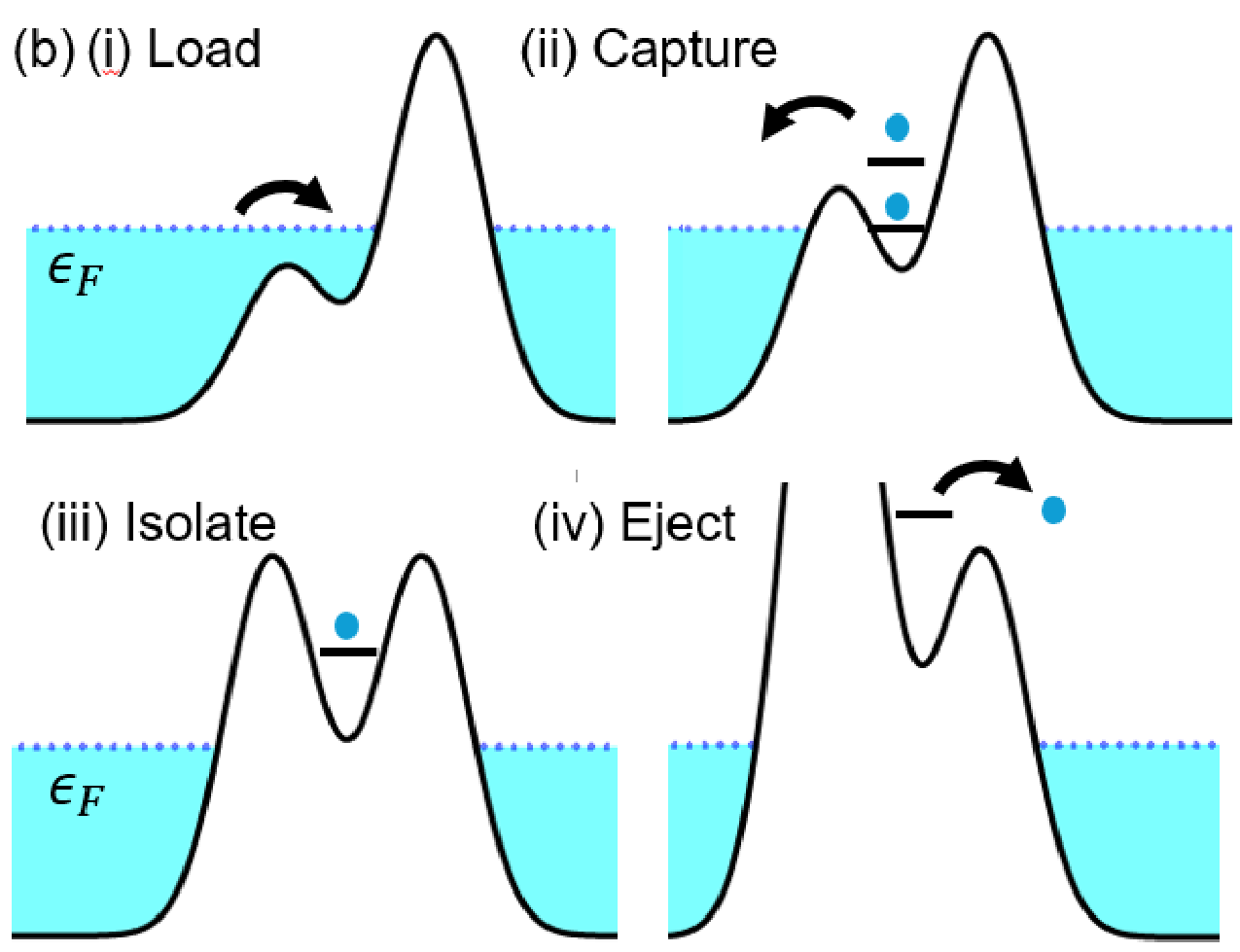}\\
	\includegraphics[scale=0.25]{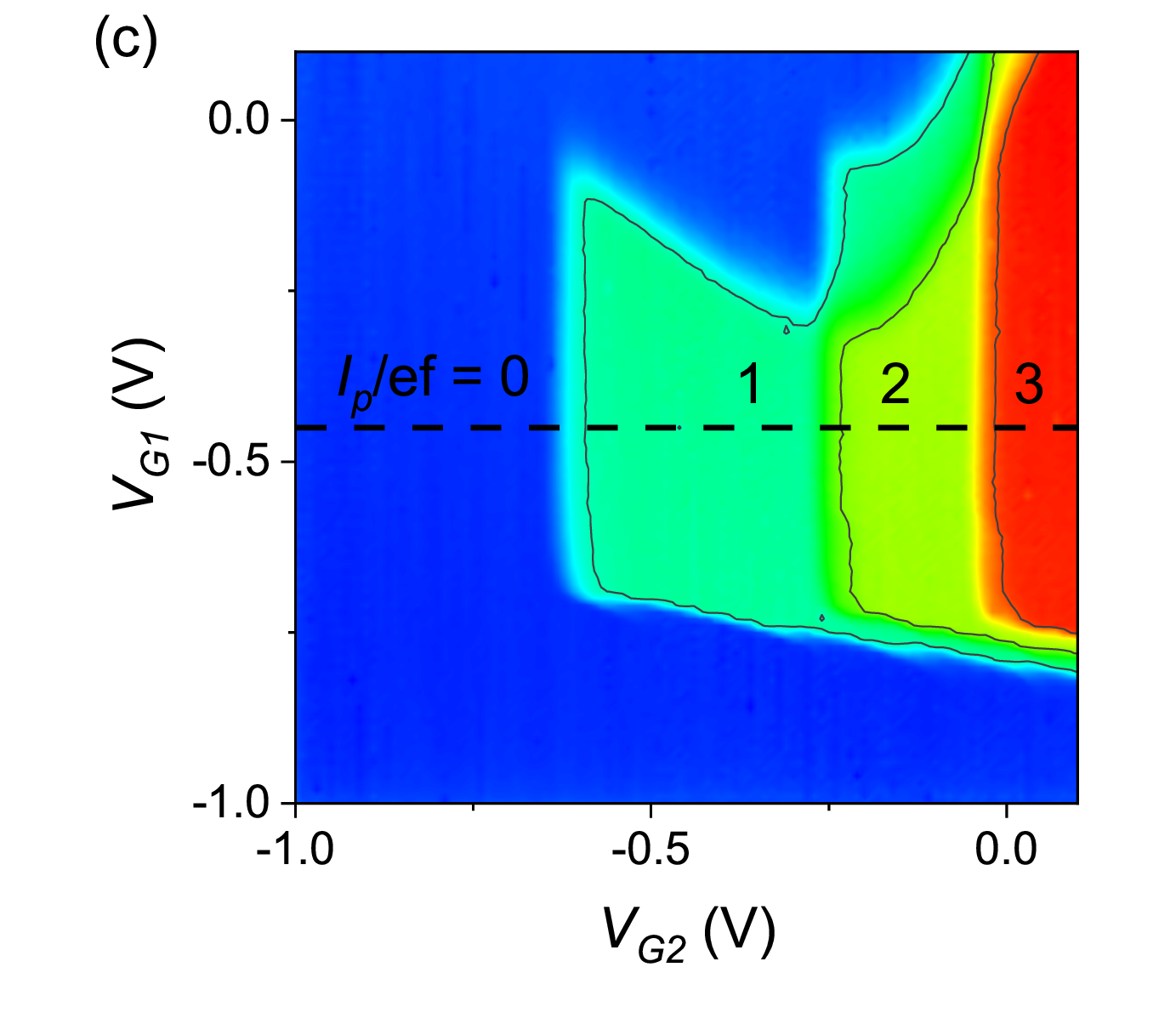}
	\includegraphics[scale=0.25]{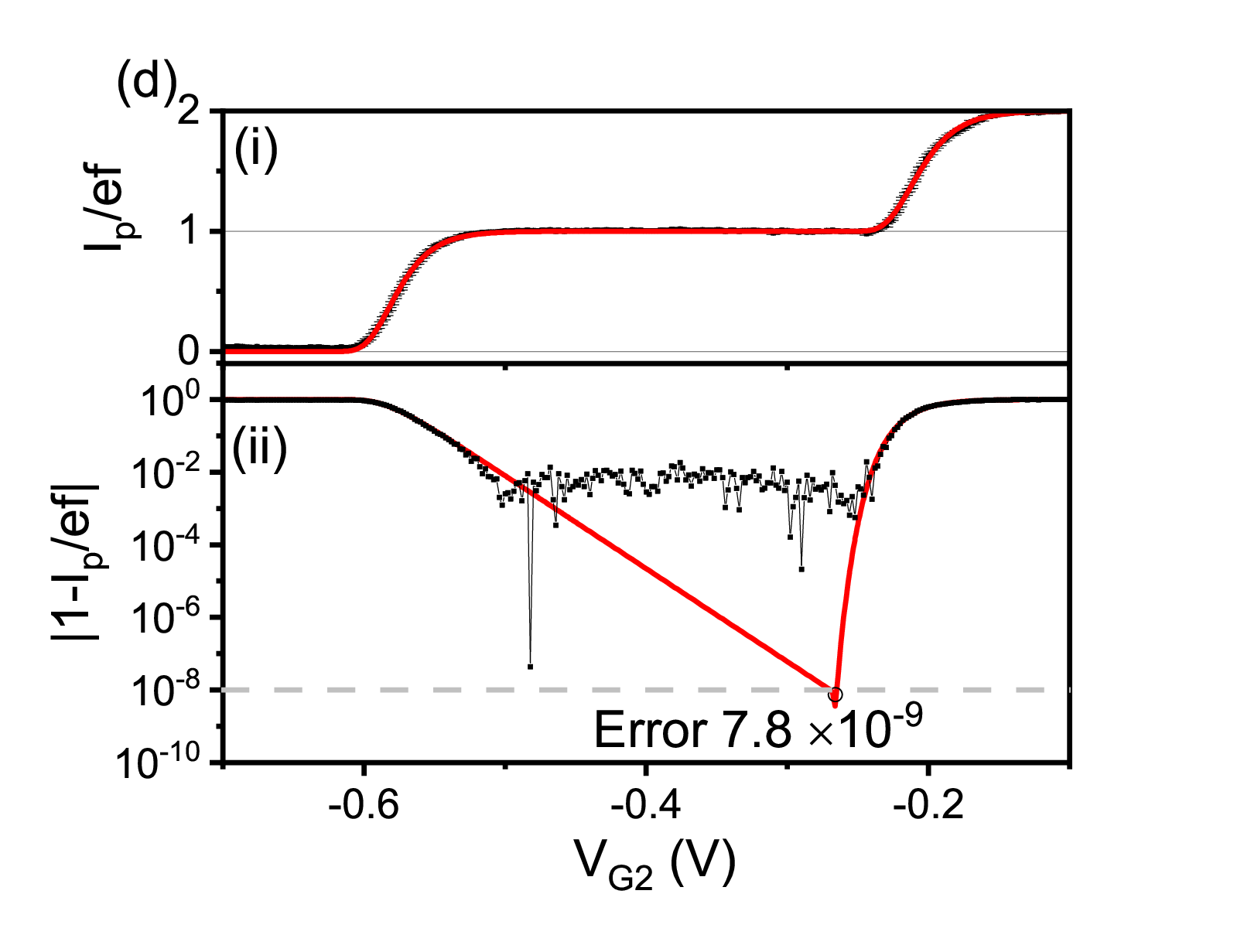}
	\caption{A single-electron pump produced on a Si MOS 300~mm process. (a) 
    Illustration of the device design and electrical connections with expected potential profile beneath.
	(b) With each cycle of $V_{\rm G1}^{\rm ac}$, charge is (i) loaded into the inter-gate QD; (ii) Sequential back-tunnelling of charge occurs, leaving capture of $n$ charges; (iii) the charge(s) are isolated in the dynamic QD; finally (iv) the charge(s) are ejected to the drain forming the pumped current $I_{p}=nef$. 
	(c) Measured current as a function of dc gate voltages, with $V_{\rm G1}^{\rm ac} \sim 0.8$~V at 100~MHz. Colour scale shows $I_{p}/ef$, denoting the number of charges $n$ pumped per cycle. Contour lines are at unit increment. 
	(d) Current measured as a function of the dc voltage on $V_{\rm G2}$ (dashed line in (c)) with a fit (red) to the decay cascade model (Eq.~\ref{dc fit}) verifying the sequential tunnelling regime of operation. 
    (ii) Fitting the deviation from single-electron tunnelling, $\left| 1-I_{p}/ef \right|$ indicates an error $<10^{-8}$. 
    }
	\label{fig:fig1}
\end{figure*}

\section{Device and Accuracy}
Figure~\ref{fig:fig1}(a) shows a sketch of our device (see Methods for further details) and the expected potential profile landscape created upon energising the gates.
A channel is defined under an accumulation gate (TG) and between two confinement gates (C1, C2) forming a path between source and drain ohmic contacts.
In the active region of the device, gates (G1, G2) define tunnel barriers in the channel, forming the QD between them. 
In regular operation, $V_{\rm C1,2} = $~0~V, and ambient temperature is $T=10$~mK.
We add a sinusoid ac voltage $V_{\rm G1}^{\rm ac}$ to fixed dc offset $V_{\rm G1}^{\rm dc}$, to drive the single-electron source at a frequency $f$.
For each period of $V_{\rm G1}^{\rm ac}$, $n$ charges (with charge $-e$) pass from  the source to the QD to the drain, producing a dc pumped current $I_{p} = nef$.

The operation of the charge pump single-electron source can be understood from the potential profile across the device at different points in the period of $V_{\rm G1}^{\rm ac}$, shown in Fig.~\ref{fig:fig1}(b). 
During loading (i), the barrier on the source side of the device is lowered to enable charges to populate the region between G1 and G2.
In the capture process (ii), a QD is formed by the rising potential barrier, while confinement in the QD causes electrons to backtunnel sequentially (the decay cascade ~\cite{Slava}) until $n$ charges remain confined in the QD.
During isolation (iii), the QD has a stable occupation $n$ and its potential rises due to cross-coupling with $V_{\rm G1}^{\rm ac}$ (discussed further below).
Finally, during ejection (iv), the QD potential is sufficiently high that the resident $n$ charges tunnel to the drain, realising $I_{p}$. 
Throughout the process, $V_{\rm G2}$ is held constant.

The current measured through the device, $I_{p}/ef$, is shown in Fig.~\ref{fig:fig1}(c) as a function of the dc gate voltages $V_{\rm G1,2}$, revealing regions of operating voltage where the pump stably produces a current for $n = 1, 2, 3$. These step-like regions are characteristic of sequential tunnelling from a dynamic QD~\cite{Slava, Kaestner3, GiblinReview, Fujiwarabook} and have been well described by the decay cascade model~ \cite{Slava}.

\begin{figure}[t!]
	\centering
	\includegraphics[scale =0.3]{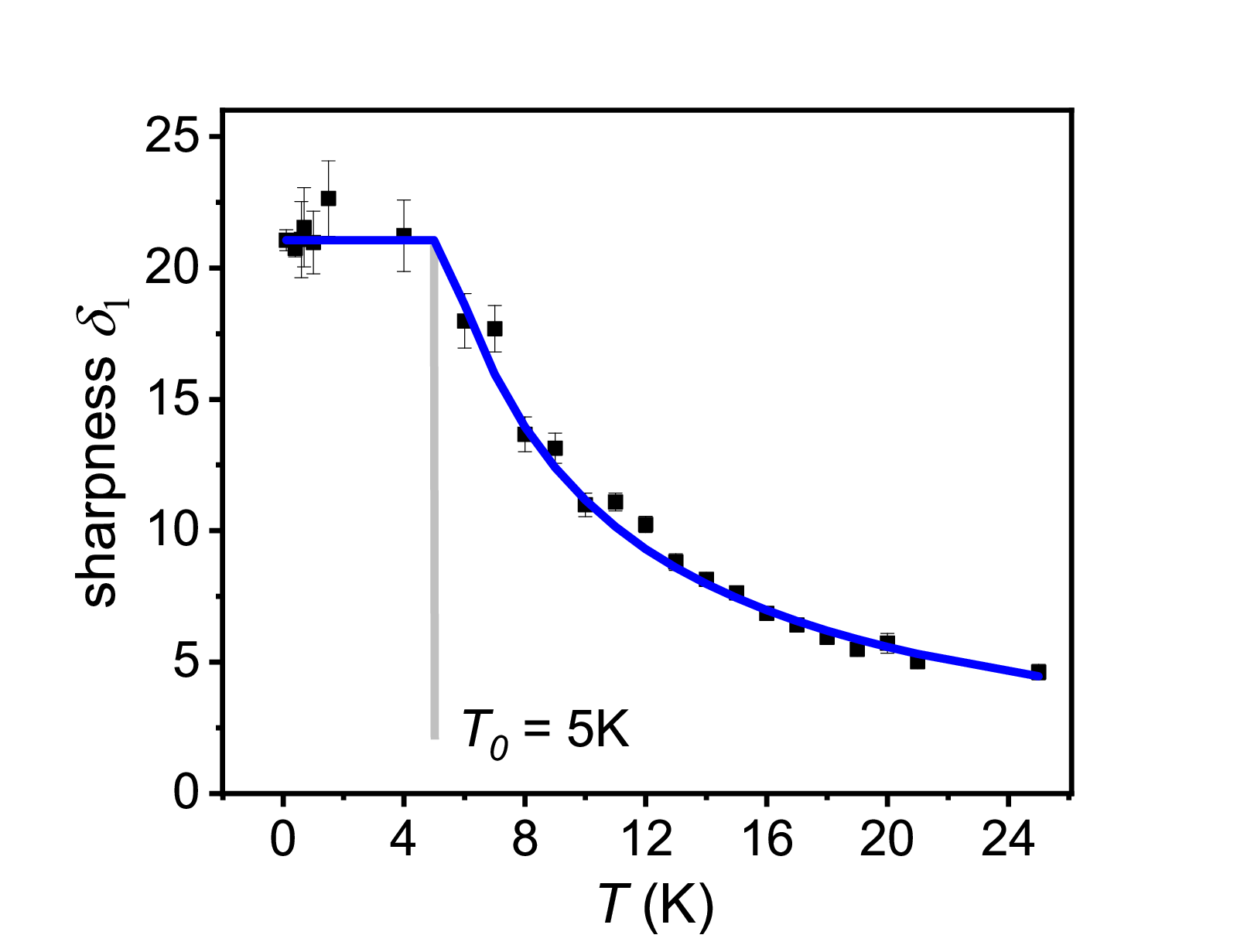}
  
	\caption{Plot of the sharpness (accuracy approximator) $\delta_{1}$ (Eq.~\ref{dc fit}) with sample temperature shows a transition at $T_{0}\sim5$~K from the tunnelling-limited regime to the thermal regime, captured by the fit to Eq.~\ref{g fit} (blue). From this we derive the properties of the QD-gate system that define the high-accuracy operation. Error bars are systematic from fit to Eq.~\ref{dc fit}.}  
	\label{fig:fig2}
\end{figure}

To evaluate the error, we focus on the case of single electron pumping ($n = 1$) illustrated in the plateau shown in Fig.~\ref{fig:fig1}(d)(i), taken at the dashed line of Fig.~\ref{fig:fig1}(c) ($V_{\rm{G1}} = -0.45$~V). In this region, the error can be expressed as $\left|1 - I_{p} / ef \right|$, as plotted in the panel (ii) (the case for higher $n$ is analogous and discussed in the Supplementary Information). 
The right-hand axis of Fig.~\ref{fig:fig1}(d)(ii) plots the (as-measured) current level $I_{p}$, and highlights the $\sim 80$~fA noise floor.
A direct measurement of the error rate is prohibited by the measurement noise floor, so we use a fit to the decay cascade model to infer the error (see for example Refs~\onlinecite{Johnson3, Yamahata3} and the Supplementary Information). An analytical solution to the decay cascade model is~\cite{Slava, Kaestner, Cole}
\footnote{The model requires the tunnel barrier formed by G1 to be smoothly varying with no local minima, for which the simplest and expected form is a parabola.}:
\begin{equation}
	\label{dc fit}
	\frac{I_{p}}{ef} =  \sum_{n = 1}^{2} \exp \left(- \exp \left(- \delta_{n} \frac{V_{\rm G2} - V_{n}}{V_{n+1}-V_{n}} \right) \right)
\end{equation}
where $V_{n}$ is the threshold of the step to the $n$th plateau.
The parameter $\delta_{n}$ is a measure of the sharpness of the current steps (and an approximator to the flatness of a current plateau) and strongly determines the accuracy of the current source. The fit to the trace in Fig.~\ref{fig:fig1}(d) gives $V_{1} = -0.581$~V, $V_{2} = -0.212$~V, $V_{3} = 0.006$~V and $\delta_{1}= 21.8$, yielding a minimum expected error of $7.8 \times 10^{-9} \pm 3 \times 10^{-9}$ for $n = 1$ (see Fig.~\ref{fig:fig1}(d)(ii)).
The error presented here is the error on the fit (see also the discussion of Fig.~\ref{fig:fig2} and the Supplementary Information).
Such a low error rate is within bounds for this device to realise a metrological definition of the Ampere and is amongst the best results reported (see Ref~\onlinecite{GiblinReview} for a review).

To understand the contributing factors to this high accuracy, captured in the high value of the sharpness $\delta_{1}$, we studied the single-electron source as a function of device temperature (see Fig.~\ref{fig:fig2}). Below 5~K, we observe a temperature-independent regime, in which quantum tunnelling through the barrier dominates,
while at higher temperatures thermal hopping reduces the pump accuracy with a characteristic $1/kT$ dependence~\cite{Johnson3, Fujiwarabook}
\footnote{We treat our dynamic QD as a memoryless system and so this elasticity removes any expectation of a small temperature dependence expected in the tunnelling limited regime \cite{Grabert2, Johnson3, Yamahata3, Yamahata5}, which is not observed at our experimental resolution.}.
The crossover temperature, $T_{0}$, at the interface between these two regimes, forms a natural figure of merit for dynamic-QD devices and can be used independently of the decay cascade model to understand the accuracy limitations.

The crossover temperature $T_{0}$ is determined by a number of coupled factors, including the potential profile of the tunnel barrier, the cross-coupling $g$ of the QD potential to gate G1 (and hence $V_{\rm G1}^{\rm ac}$),
and the charging energy of the QD, $E_{\rm c}$.
The cross-coupling $g$ can be expressed as $g = \alpha_{\rm G1,QD}/\left(\alpha_{\rm G1,barrier}-\alpha_{\rm{G1,QD}}\right)$ where $\alpha_{\rm{G1,QD}}$ and $\alpha_{\rm G1,barrier}$ are the gate lever arms between G1 and the QD and barrier potentials.
Following the same approach as Ref [\onlinecite{Johnson3}], we fit the temperature dependence of $\delta_{1}$ to
\begin{equation}
\delta_{1} = \left( 1 + \frac{1}{g} \right) \frac{E_{c}}{k_{B}T^{*}} 
\label{g fit}
\end{equation}
with $T^{*} = T$,  when device temperature (cryostat) $T > T_{0}$ or $T^{*} = T_{0}$ when $T \leqslant T_{0}$, and $k_{B}$ Boltzmann's constant.
From the fit, we obtain $T_{0} = 5.0(5)$~K and $E_{c} = 8(1)$~meV, with uncertainties determined by the method presented in the Supplementary Information.
Previous results have reported $T_{0}$ around 10~-~20~K \cite{Johnson3}, making our result one of the best in class under this assessment, which in turn supports the state-of-the-art charge pump fidelity reported earlier in this manuscript.
We see that Eq.~\ref{g fit} demonstrates independently of Eq.~\ref{dc fit} the requirements to make a high accuracy device. 
It follows that we require a high charging energy, $E_{c}$, to minimise the error from multiple occupation of the QD, and that $T_{0}$ should be as low as possible to achieve the highest value of $\delta_{1}$, by intercepting the thermal decay curve at a higher value of $\delta_{1}$.
The fitted value of $E_{c}$ is dependent on the choice of value for $g$, and we have assumed $g \geq 5$ in the fit described above. This assumption is based upon the observation of the decay cascade model remaining a good fit throughout the temperature range studied here~\cite{Johnson3} and is discussed further in the Supplementary Information. 

For an elastic QD, we can express the cross over temperature determined above to properties of the quantum dot and tunnel barrier: $T_{0} = \hbar \omega_{\rm G1}/2k_{B}\pi$, where $\omega_{\rm G1} = -\frac{1}{m^{*}} \frac{{d}^{2}}{{d}x^{2}}$~$V(x)$ is the G1 barrier frequency; $m^{*}$ the electron reduced mass and $V(x)$ the potential of the barrier in real space $x$ \cite{Hanggi}. From a $T_{0}$ of 5~K we extract $\omega_{\rm G1} = 4.1$~THz. Taking a quadratic $V(x)$ and a barrier height of G1 at the time of capture (Fig.~\ref{fig:fig1}(b)(ii)) as slightly less than $E_{c}$, we write $E_{c} = m^{*} \omega_{G1}^{2} x^{2} / 2$.
Taking $E_{c}$ as an upper limit, we find $x \approxeq 30$~nm, which is consistent with the G1 lithographic gate length of 40~nm (which is reduced after oxide growth).

In the absence of a direct measurement of the error rate in our device, these consistent observations in the single-electron dynamics that govern the charge transfer process provide confidence in the decay cascade model as a good representation of the single-particle dynamics of the pump, and therefore a fit to that model (through Eq.~\ref{dc fit}) as a good estimate of its underlying accuracy.

\section{Analysis across the two-dimensional Parameter Space}

\begin{figure}[ht!]
	\centering
	\includegraphics[scale=0.3]{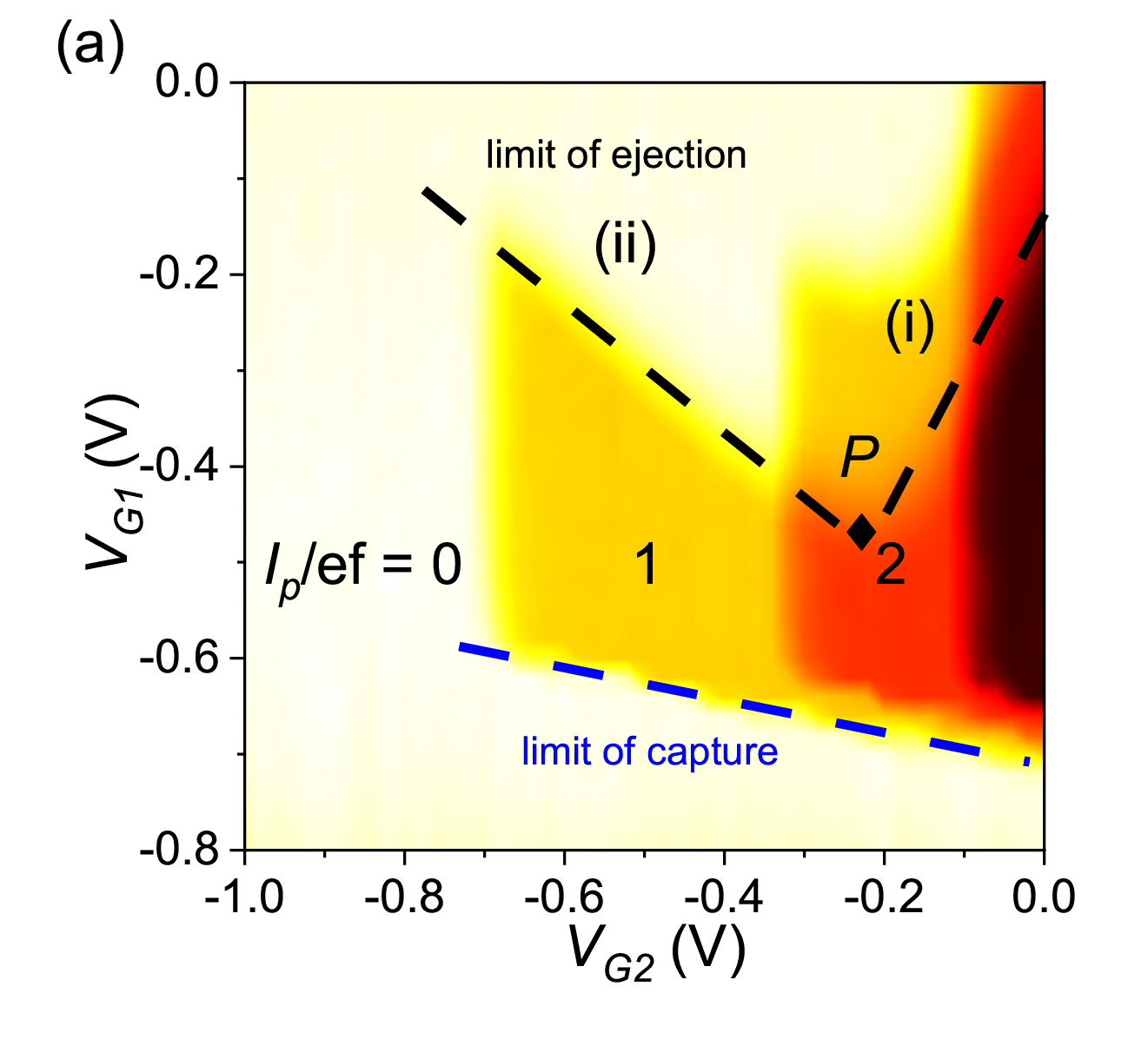}
	\includegraphics[scale=0.3]{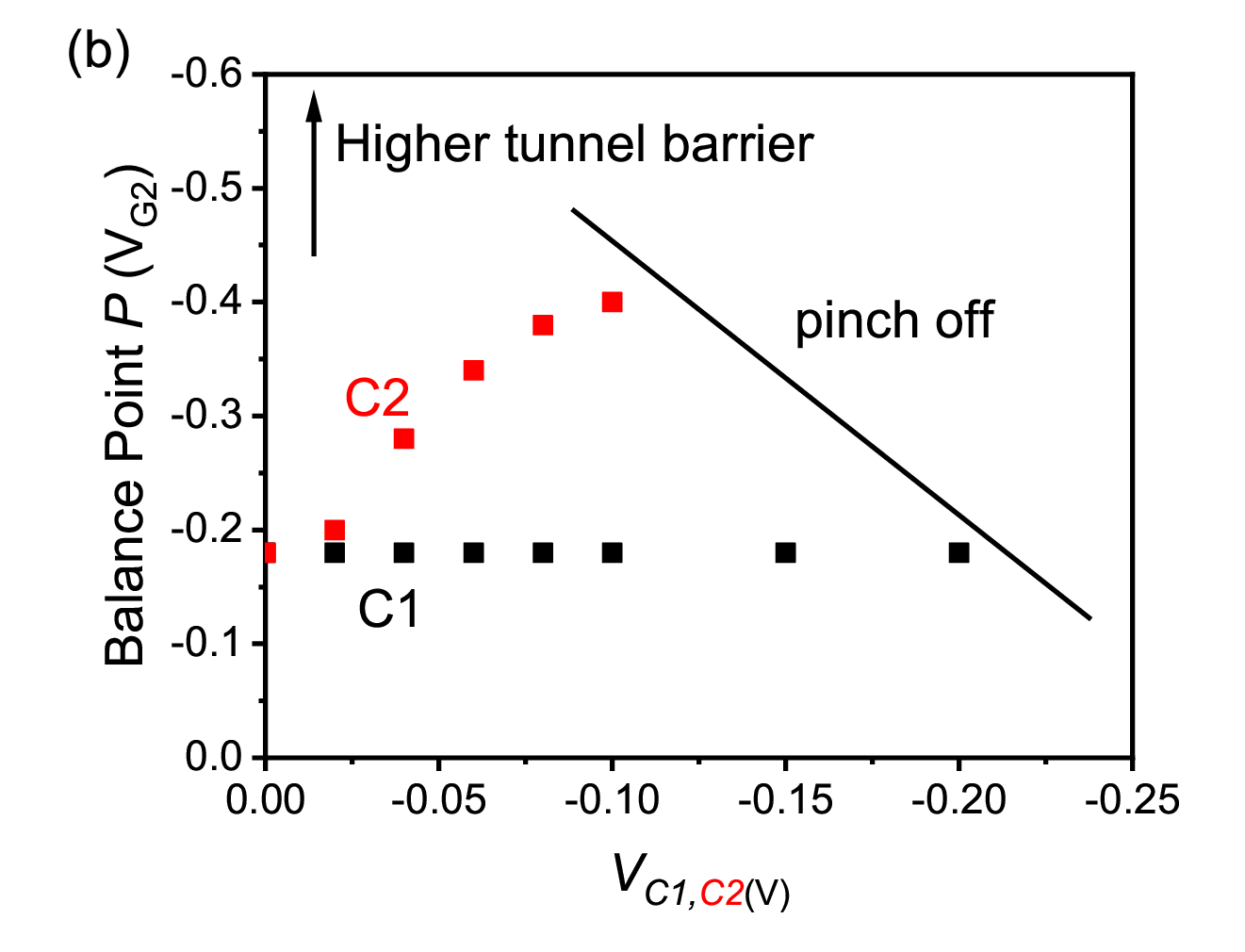}
	\includegraphics[scale=0.3]{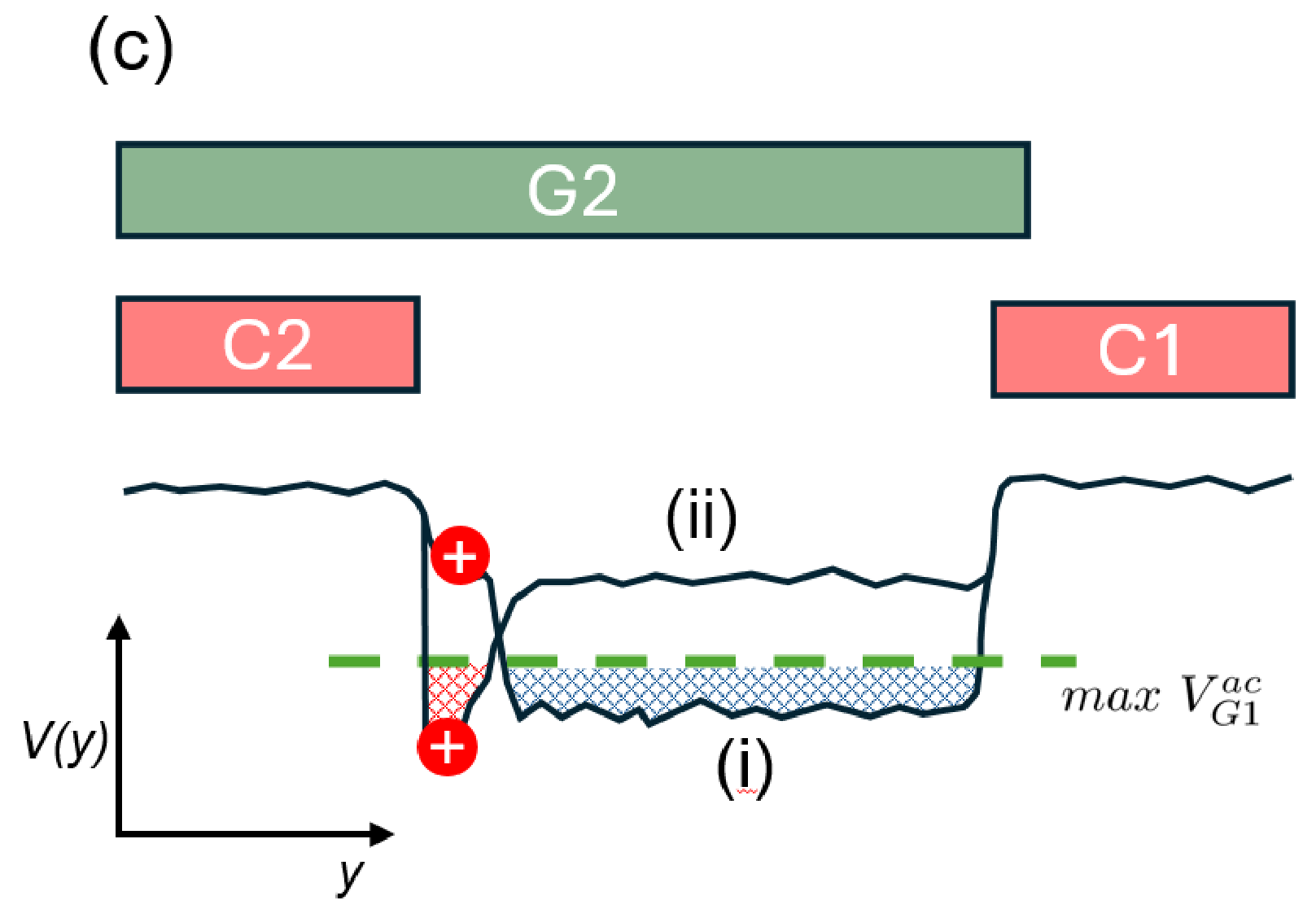}
	\caption{(a) 2D map, analogous to Fig. 1(c), but at $V_{\rm TG} = 1.1$~V, shows clear presence of additional bounding line (ii), contrary to the only expected loading bound (i). (b) The vertex between the split loading lines – the balance point $P$ - is seen to vary in $V_{\rm G2}$  with $V_{\rm C2}$ but not $V_{\rm C1}$. (c) We derive the balance point $P$ as the point at which the ejection tunnel barrier height aligns with the QD energy at the time of  maximum potential (stage (iv) in Fig. \ref{fig:fig1}(c), towards the maximum value of $V_{\rm G1}^{\rm ac}$). For lower barriers, case (i), ejection can occur across the G2 span. For higher barriers, case (ii), ejection only happens in the pinned state for values $V_{\rm G1} >$ case (i).
	}
	\label{fig:fig3}
\end{figure}

To gain further understanding on the origin of the high inferred pump accuracy, we now turn our attention to the 2D parameter space spanned by the gate voltages $V_{\rm G1}$ and $V_{\rm G2}$.
Figure~\ref{fig:fig3}(a) plots the pumped current $I_{p}/ef$ in the colour scale, taken at $f = 100$~MHz.
The current map indicates the same overall bounding conditions as reported in this class of electron pump \cite{Kaestner3}, and trapezoidal regions of operation similar in shape to those predicted from sequential single-electron tunnelling~\cite{GiblinReview, Fujiwarabook} (see Supplementary Information for more detailed mappings). 
The region of quantised current is bound by the limits of the difference between the QD potential, the G1 and G2 barrier potentials and the charging energy $E_{c}$. The lower bounding line corresponds to the limit of capture (and loading), as described in stage (ii) of Fig.~\ref{fig:fig1}(b). 
For values of $V_{\rm G1}^{\rm dc}$ above this value (lower potential), the dynamics proceed according to Fig.~\ref{fig:fig1}(b).
The upper bounding line corresponds to the limit of ejection (stage (iv) of Fig.~\ref{fig:fig1}(b))).
At more positive voltages above this bound, the QD potential does not rise sufficiently to allow escape to the drain and the electron is returned to source.
In our device, the limit of ejection undergoes a change in gradient, with the conventional limit marked as (i) and the anomalous limit as (ii) in Fig.~\ref{fig:fig3}(a)).

We can understand the additional region of pumping, and the bound (ii), by considering the presence of a positive charge trap.
The role of the positive charge trap can be examined by studying the role of the channel confining gates C1 and C2. 
In our discussion so far, these have been held at $V_{\rm C1,2} =0$~V to allow the conductance path to be defined entirely by $V_{\rm TG}$.
In Fig.~\ref{fig:fig3}(a), we define a balance point $P$ as the intersection of ejection lines (i) and (ii), and in Fig.~\ref{fig:fig3}(b) we plot the position of $P$ with respect to $V_{\rm G2}$, for each of $V_{\rm C1,2}$, with the other gate held at 0~V.
We see that $P$ is not seen to move across the map of Fig.~\ref{fig:fig3}(a) with $V_{\rm C1}$, but is linearly dependent on $V_{\rm C2}$.
This capacitative coupling allows us to infer the presence of a positive charge trap that mediates tunnelling to drain.
Figure~\ref{fig:fig3}(c) shows a sketch of the potential profile under G2, as a cross-section across the channel.
We omit TG as it is held constant throughout this measurement.
Due to the strong dependence on $V_{\rm C2}$ and not $V_{\rm C1}$, we suggest the positive trap state exists around the G1-C2 interface at the edge of the conducting channel.
The line denoted $max\left(V_{\rm G1}^{\rm ac}\right)$ on Fig.~\ref{fig:fig3}(c) shows the maximum potential of the QD at its highest point in the period $1/f$; this determines the position of the bounding lines (i) and (ii).
Tunnelling will hence occur depending on the relative barrier height of G2 to this level (condition (i)) or the trap state (condition (ii)).
With reference to Fig.~\ref{fig:fig3}(c), in the conventional, barrier-mediated region (i), the barrier height $V_{\rm G2}$ is close to or lower than $max\left(V_{\rm G1}^{ac}\right)$ allowing escape via G2 to form $I_{p}$ (blue hatched region). 
In condition (ii), $V_{\rm G2} > max\left(V_{\rm G1}^{\rm ac}\right)$ which would normally not permit any current flow.
However, we see the positive trap state, indicated on Fig.~\ref{fig:fig3}(c), can still mediate charge escape.
The positive trap  is capacitively coupled to $V_{\rm G2}$ which forms the dependence seen in condition (ii) (red hatched region) .
The charge trap is seen to be more strongly capacitively coupled to $V_{\rm C2}$ than $V_{\rm G2}$, which allows $V_{\rm C2}$ to shift $P$.

The contiguity of the observed limits suggest that electrons always load into the quantum dot and the role of the trap is only to enlarge the region where electrons are ejected from the QD. A trap-mediated current existing in parallel with our QD would result in a second pumping map that would superimpose, and we would not expect to see the continuity of plateaus as we cross $P$.
By providing a greater range of $V_{\rm G2}$ for which ejection can occur, the trap enlarges the $n = 1$ plateau, which contributes to the high pump accuracy observed (in addition to the high sharpness and crossover temperature discussed above for barrier G1).
While the presence of a trap is not necessary in principle to construct high accuracy devices, the observed behaviour motivates considering device modifications, for example in the construction of smaller QDs or barriers to replicate the effect of this trap.

Further, we note that there is no loss of accuracy across any of the region bounded by both conditions (i) and (ii), and as we apply $V_{\rm C1,C2}$ (Fig.~\ref{fig:fig3}(a) and Supplementary Information for a quantitative study). 
This shows that gates $\rm C_{1,2}$ can be used as an energy selector for the emission energy of electrons from the pump: they simply move the position of the pump map via capacitance effects, and do not play a role in determining the accuracy, which as we have shown is determined by the form of barrier $\rm G_{1}$.
This is in contrast to previous results \cite{Rossi} where side gates have played a role by tuning the QD confinement, which may impede scale-up by requiring more tuning. 
This unchanging accuracy verifies the mechanism of operation previously presented, and serves to show we can incorporate some level of imperfection in the nano-fabrication and maintain accuracy.

\section{Discussion}
We have developed and characterised a highly accurate dynamic-QD based single electron source fabricated in a silicon 300~mm process. 
The industrial fabrication process and the robustness to device imperfections encourages the scaling to a large number of devices to achieve technologically useful nanoampere levels of current~\cite{Dash, Yamahata7}.
One approach to assist with scaling is to move from independently-controllable devices to ones where gates are shared across multiple devices~\cite{Dash, Yamahata7}, however, this can result in many devices operating in regimes away from the point of minimum error.
Based on the results above, we can consider shared barrier gates G1 and G2 (including the AC component), combined with independently controlled C1 and C2 to tune each device to an optimum point. The capacitative effect on the position of $I_{p}$ in $V_{\rm C1, C2}$ gate-space (see Fig.~\ref{fig:fig3} and Supplementary Information) would allow, in principle, many more devices to align their points of highest accuracy with a common $V_{\rm G2}$, removing some of the tuning challenges in previous shared-gate demonstrations~\cite{Yamahata7}.

To further improve this class of device, we suggest making the QD smaller - this will keep $E_{c}$ large and replace the role of the impurity. By use of a common drive $V_{G1}^{ac}$ and independent $V_{C}$, many devices can be tuned to operate in parallel, despite local QD imperfections, vastly increasing the yield.  

In principle, a device such as the one studied here could make a key contribution to closing the Quantum Metrological Triangle (QMT) on a single silicon chip \cite{Keller3, Scherer, Scherer2, Stock}.
This proposed experiment, which aims to perform an all-in-one determination of the Ampere, Ohm and Volt, is challenging due to the inaccuracy of single-electron sources combined with the difficulty in connecting the three legs of the measurement.
Josephson Junctions patterned on silicon wafers have been used to realise a Voltage standard~\cite{Rufenacht} and could be combined with the SiMOS single-electron current source we have demonstrated. The remaining arm of the triangle, the resistance standard, could be achieved using the anomalous quantum Hall effect~\cite{Okazaki}, for example by patterning  thin-film quantum Hall materials on silicon, in a similar process to that used to define micromagnets for electron spin driving~\cite{Dehollain, McNeil}.

\section*{Acknowledgements}
We thank Andrew Fisher for helpful discussions.
This work received support from from the Engineering and Physical Sciences Research Council (EPSRC) through the Hub in Quantum Computing and Simulation (Grant No.\ EP/T001062/1).

\section*{Author Contributions}
N.J. conceived of the experiment, conducted the measurements and analysis. N.J. and J.J.L.M. wrote the manuscript with input from M.F.G.-Z. and R.C.C.L.
S.K., J.J., Y.C., led by K.d.G., developed the 300-mm silicon MOS process and fabricated the device.
J.J.L.M. provided supervision and oversight.

\section*{Methods}
\subsection{Device Fabrication}
The device is fabricated on an isotopically purified Si-28 wafer; although we importantly note this purification is not necessary for the operation of our pump, and likely plays no role in its accuracy, as this is entirely electrostatically determined.
The fabrication proceeds similarly to that detailed in Ref.~\onlinecite{Elsayed}.
In our case, after wafer growth, 20~nm of silicon oxide (thermally annealed) is formed before electron beam lithography of poly-Si gates.
Gates are written in three layers (see Fig.~\ref{fig:fig1}(a)): firstly $\rm C_{1,2}$, then $\rm G_{1,2}$ and finally $\rm G_{TG}$.
Between the first and second, and second and third layers approximately 6~nm of oxide is formed (via deposition).
Optical lithography is employed for the larger poly-Si structures away from the active regions of the device. 

\subsection{Measurements}
The sample is cooled to a base temperature of 10~mK in a Bluefors XLD dilution refrigerator at zero magnetic field. 
dc voltages are supplied by a QDevil QDAC Mk.I, with $V_{\rm G1}^{\rm ac}$ supplied by either an Agilent 8257C or Rohde \& Schwarz SMC100A, with an output power of 9~dBm which is approximately 4-5~dBm on chip.
dc signals are passed through a low-pass filter (cut-off 16kHz).
A resistive bias tee is used to combine $V_{\rm G1}^{\rm ac}$ with $V_{\rm G1}$.
A custom-made PCB containing the dc filtering and bias tee is used to house the chip; connections are made with Al-1\%Si wirebonds.
$I_{p}$ is measured directly with a Spectrum M4i.4421-x8 digitiser; a Stanford Research SR570 provides I/V conversion and gain.
For the accuracy evaluation, we take an average of 10 repeat traces in $V_{\rm G2}$  (6 for plots in the Supplementary Information).
For the temperature dependent measurements of Fig.~\ref{fig:fig2}, we make use of both the mixing chamber heater and heat-switches to stabilise the temperature; the pulse-tube cooler is left on.
Expected drift in temperature is maximal at lower temperatures and should not exceed $\sim~0.2$~K.

\section{Supplementary Information}
\renewcommand\thefigure{S\arabic{figure}} 
\setcounter{figure}{0}

\subsection{Accuracy with $V_{\rm TG}$}
\begin{figure*}
    [h!t]
	\centering
	\includegraphics[scale=0.25]{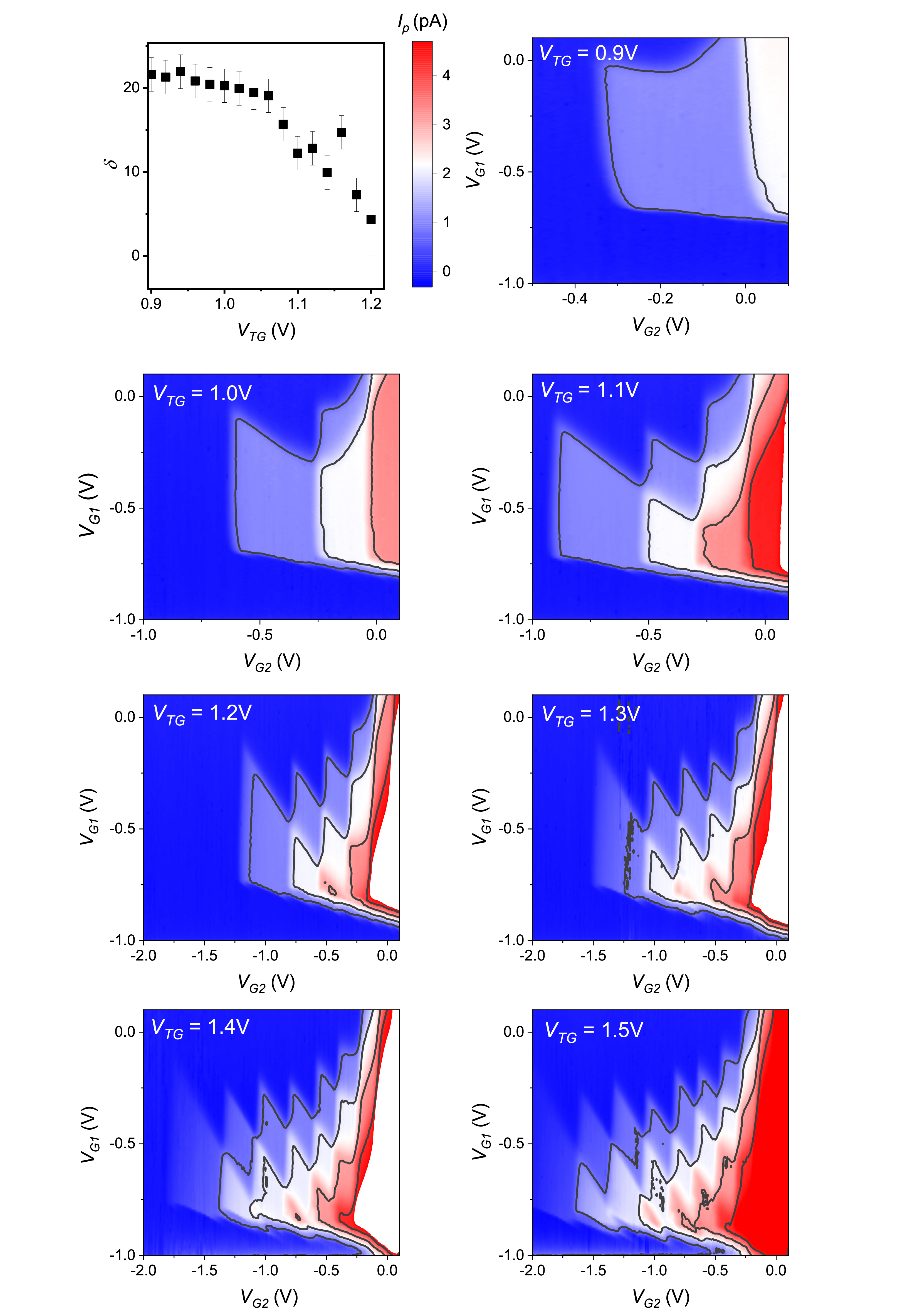}
	\caption{Pump maps with varying $V_{\rm TG}$. Contours mark successive electrons pumped per cycle. The breakdown in the single electron tunnelling regime is seen with increasing $V_{\rm TG}$.  }
	\label{fig:S1_TG}
\end{figure*}
Fig.~\ref{fig:S1_TG} plots the evolution of the pump map with $V_{\rm TG}$, taken at $f = 100$~MHz and $V_{\rm C1} = V_{\rm C2} = 0$~V.
Fig~\ref{fig:S1_TG}(a) plots the sharpness $\delta_{1}$, found from a fit to eqn.~\ref{dc fit} across the plateau, with error as the error to the fit to eqn.~\ref{dc fit}.
We note plots in Figs.~\ref{fig:fig1} and \ref{fig:fig2} are taken at $V_{\rm TG} = 1.0$~V. 
In the colour maps, contour lines mark successive electron number $n$ per cycle.
We see that the accuracy degrades with increasing $V_{\rm TG}$, and by $V_{\rm TG} \approx 1.3$~V, we find the decay cascade process to have broken down, and the characteristic double-exponential step-like progression between plateaus (see Fig.~\ref{fig:fig1}(c)) has become more linear.
We hypothesise that at these higher values of $V_{\rm TG}$, the single-electron tunnelling events break down as either the second principal charge mode becomes populated in the lead \cite{Tilke} and/or potential fluctuations under $\rm G_{1}$ allow multiple tunnel paths.
Interestingly, despite the first plateau showing this transition from single to multiple tunnelling events, including in the higher order first plateaus (retained occupation of charge per cycle) the second plateau remains robustly decay cascade, as the marked contour line shows, potentially increasing the range of $V_{\rm TG}$ this device works at.

\FloatBarrier
\subsection{Accuracy with $V_{\rm C1,C2}$}
\begin{figure}[ht]
	\centering
	\includegraphics[width=1\linewidth]{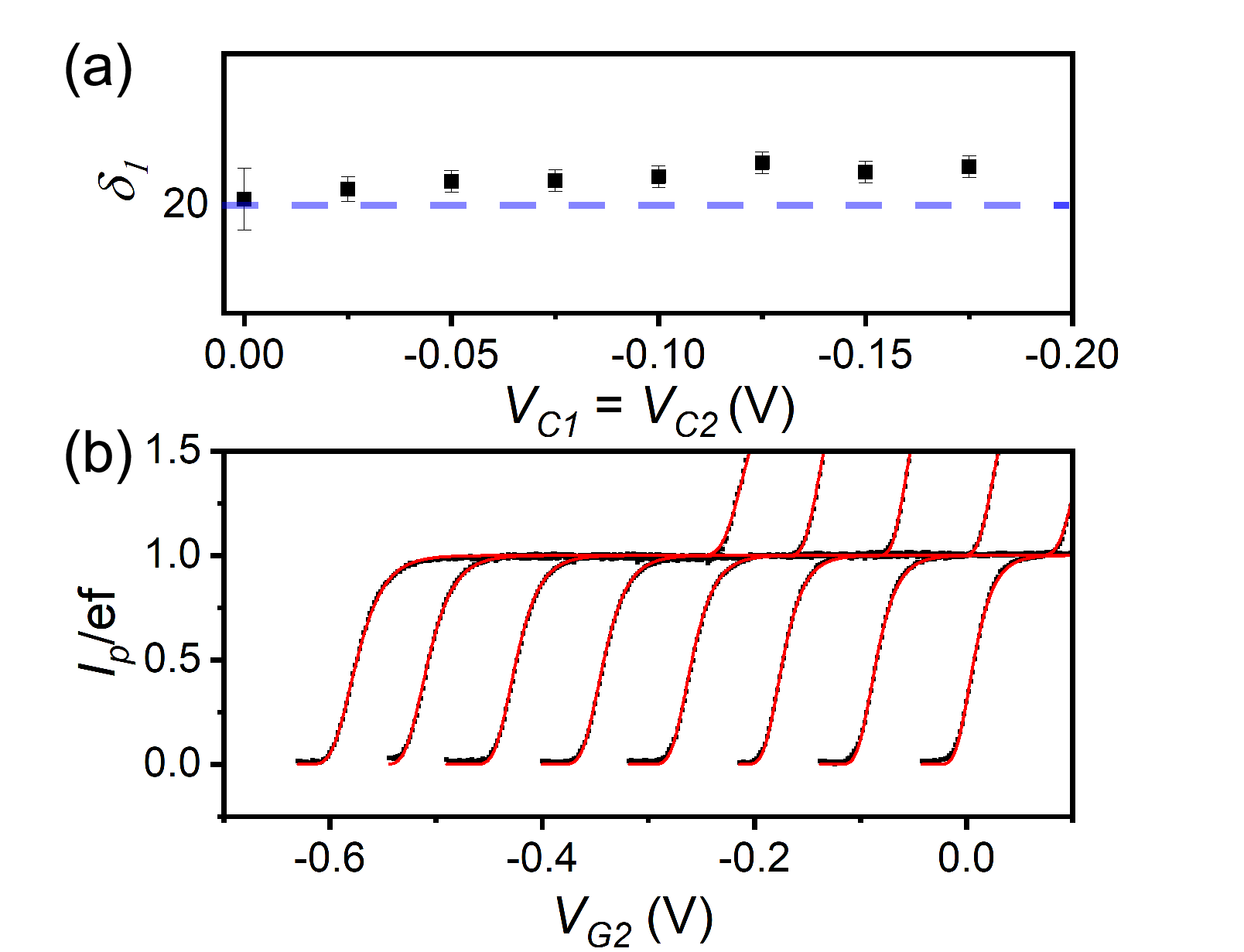}
		\caption{Accuracy evaluation under the effect of the C gates. (a) the sharpness $\delta_{1}$ from the fits with varying C-gate voltage shows no trend. (b) data with a fit to eqn.~\ref{dc fit}.}
	\label{fig:S_Cplot}
\end{figure}

Fig.~\ref{fig:S_Cplot} plots a cut across the first electron plateau at $V_{\rm TG} = 1$~V, and $f = 100$~MHz for varying $V_{\rm C1} = V_{\rm C2}$.
In Fig.~\ref{fig:S_Cplot}(b), we plot the raw data traces, and perform a fit to the decay cascade model, eqn.~\ref{dc fit}, to evaluate the accuracy (red).
In Fig.~\ref{fig:S_Cplot}(a), we plot the sharpness $\delta_{1}$ from the fit to eqn.~\ref{dc fit}, with the error as the error on the fit to eqn.~\ref{dc fit}.
The dashed line marks $\delta_{1} = 20$, which is close to the value found in Figs.~\ref{fig:fig1}(d)(ii) and \ref{fig:S1_TG} (see Supplementary Information A).
This shows us the accuracy is unchanged under the effect of the C gates, as we would expect under $\rm G_{1}$ dominating the decay cascade.
There is a positional shift in the plateaus, due to the capacitative coupling of the C gates to the QD and other barriers.
Unlike the positive charge trap explored in Fig.~\ref{fig:fig3} of the main text, which we consider a quirk of this particular device, this general effect shows us that this gate stack and geometry is generally effective.
Further, for applications where we may want energy selection, such as pipeline qubits or quantum sensing \cite{Sofia}, we see the C gates can effectively change the energy of electron emission from the pump with no loss of accuracy.

\FloatBarrier
\subsection{Accuracy evaluation for higher plateaus}
It is unusual for pumping papers to consider higher electron occupations per cycle, as these higher plateaus are universally more erroneous than the first.
However, we include some analysis here as a starting point for the community. 
Not only is pumping more than a single charge per cycle a route to scalability, but it may become crucial in future electron sources - an entangled pair pump, for example; or interferometry between wavepackets. 
So here we present a first analysis, analogous to Fig.~\ref{fig:fig1}(d), and encourage future pump developers to examine their higher electron number plateaus also.
For this, we choose the slightly higher $V_{\rm TG} = 1.1$~V, compared to $V_{\rm TG} = 1$~V used in the accuracy determination of Fig.~\ref{fig:fig1}(d). 
As we see in the Supplementary Information A, the accuracy at this value is the same, and it provides a reasonable number of plateaus to evaluate before the breakdown to the decay cascade limit.

\begin{figure}[ht]
	\centering
	\includegraphics[width=1\linewidth]{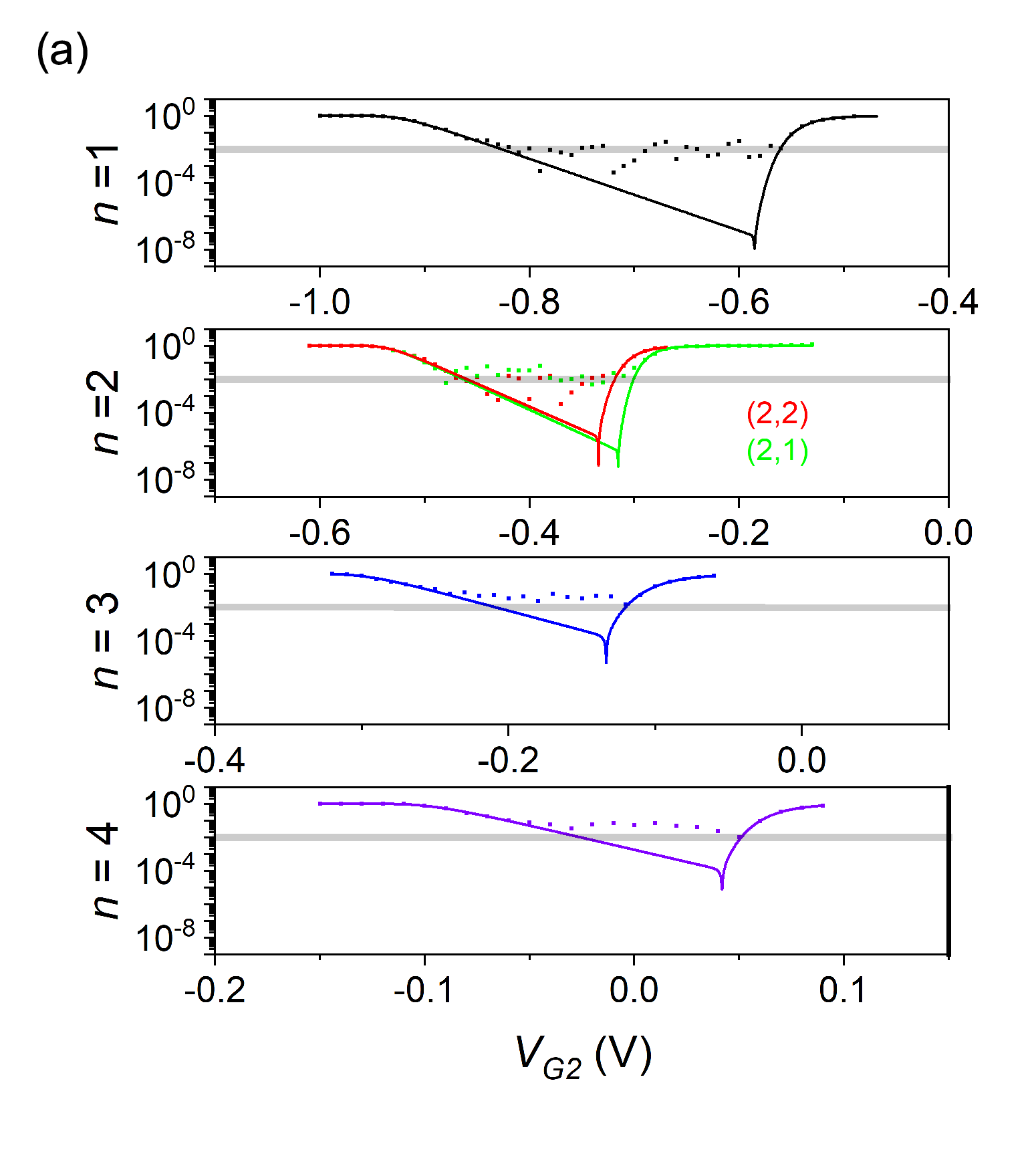}
	\includegraphics[width=1\linewidth]{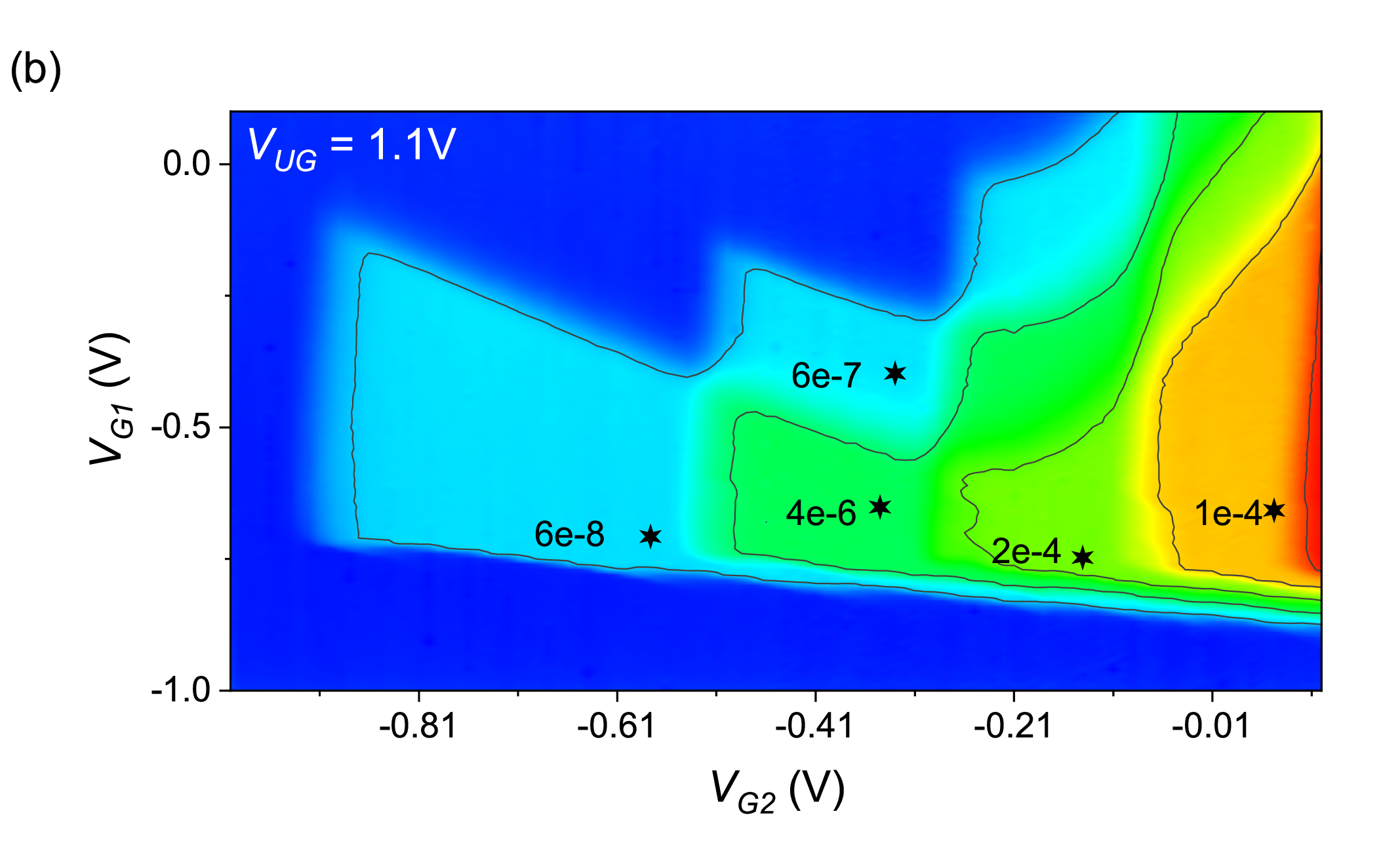}
	\caption{(a) Plots of $\left| n - I_{p}/ef \right|$ for successive $n$, with fits to the decay cascade model, eqn.~\ref{dc fit}. The noise floor is highlighted with a grey line. Contour lines denote threshold for successive $n$.
	(b) The full pumping map for multiple $n$, with locations of the best accuracy determined marked and recorded.
	}
	\label{fig:S_nplot}
\end{figure}

Fig.~\ref{fig:S_nplot}(a) presents accuracy analogously to Fig.~\ref{fig:fig1}(d), for various plateaus across the pump map.
In each plot, the vertical axis plots $\left| n-I_{p}/ef \right|$, and the horizontal $V_{\rm G2}$. 
Each plot in (a) is indexed by the number of charges pumped per cycle, $n$.
To understand the map, we break down $n$ further into the coordinate pair of (number loaded, number ejected).
On the principal plateaus, these are equal. 
For $n > 1$, we see this plateau splits into several plateaus with incremental numbers of ejected electrons per cycle.

In each plot, the gray line represents the noise floor ($10^{-3} ef \approx 20$~fA).
We should not expect that the accuracy as determined by the fit to eqn.~\ref{dc fit} to be experimentally obtainable for $n > 1$, as there may be other underlying multi-particle physics that comes into play in a multiply-occupied QD.
However, as we see in Fig.\ref{fig:S_nplot}(a), for all cases here the fit to eqn.~\ref{dc fit}, and hence the validity of the decay cascade model, is good at the inter-plateau steps.
We note that the $n = 4$ plateau is slightly larger and of better quality than $n = 3$.
This may be due to the complete shell structure of a four electron quantum dot in silicon.

\FloatBarrier
\subsection{Parametrising $g$}
\label{SD}

In this section, we justify our use of selecting $g=5$, as stated in the main text.

The cross-coupling $g$ can be expressed as $g = \alpha_{\rm G1,QD}/\left(\alpha_{\rm G1,barrier}-\alpha_{\rm{G1,QD}}\right)$ where $\alpha_{\rm{G1,QD}}$ and $\alpha_{\rm G1,barrier}$ are the gate lever arms between $\rm{G_{1}}$ and the QD and barrier potentials.
We note that $g = 0$ is thermal equilibrium (no QD movement with respect to the lead), and $g > 0$ describes more coupling of the QD to $V_{\rm G1}^{\rm ac}$, and therefore more movement of the QD.
It is seen that this parameter is of key importance in the accuracy of our device \cite{Yamahata3, Yamahata6}. 
This is because $g$ will determine the energy of the QD with respect to the lead and barrier $\rm G1$, which determines the back-tunnelling rate at the time of decoupling from the lead, independently of the decay cascade model.

To parameterise $g$, we make the following three arguments.
Firstly, we see from Fig.~\ref{fig:fig2} that a clear cross-over temperature $T_{0}$ exists. 
This strongly suggests the condition $g > 0$, i.e. the error observed below $T_{0}$ is not due only to thermal hopping \cite{Yamahata3}. 
Next, we evaluate the second derivative of the pumped current, $d^{2}I_{p}/dV_{G2}^{2}$, and examine the ratio of the positive and negative amplitudes of this form.

\begin{figure}[h!]
	\centering
	\includegraphics[width=1\linewidth]{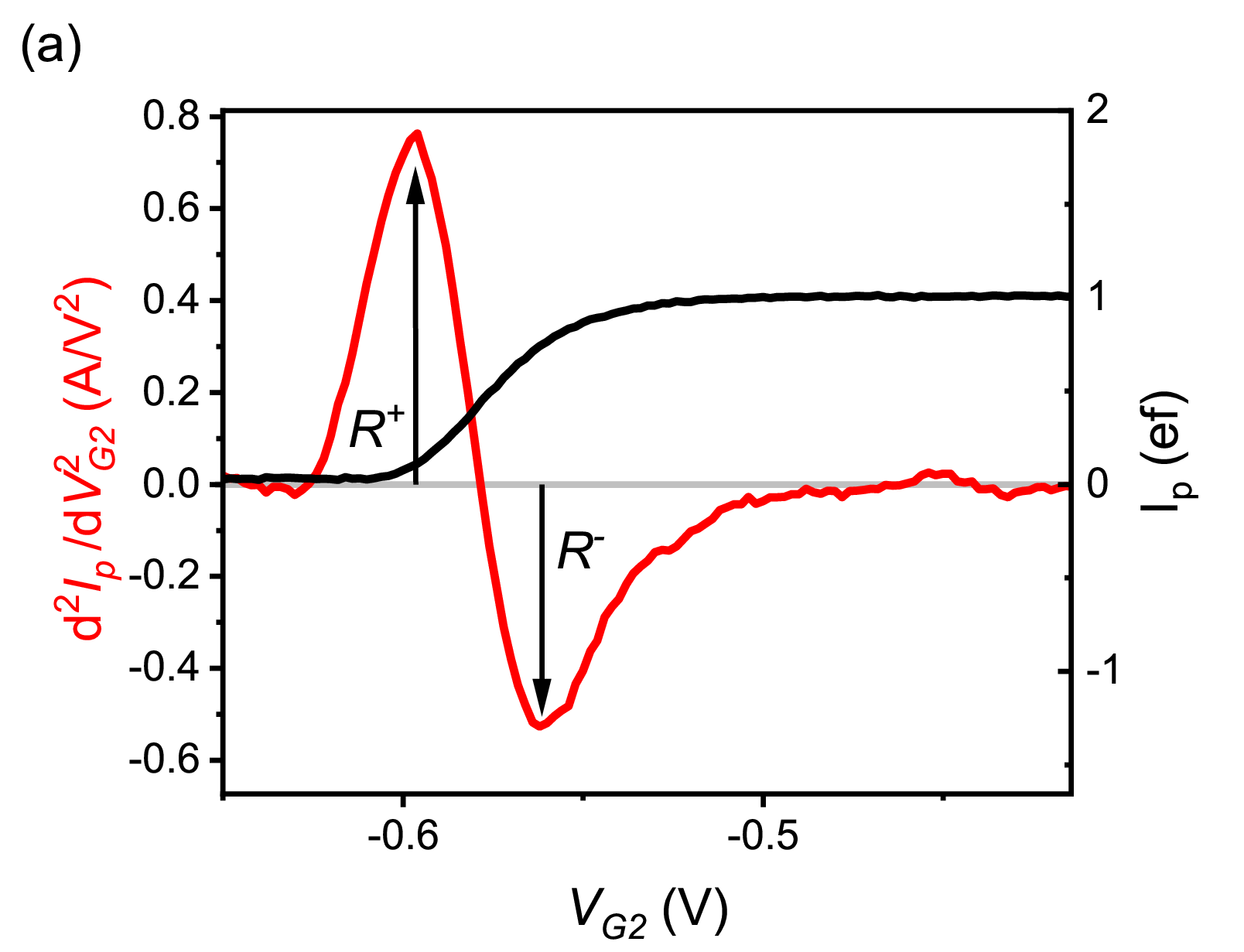}
	\includegraphics[width=1\linewidth]{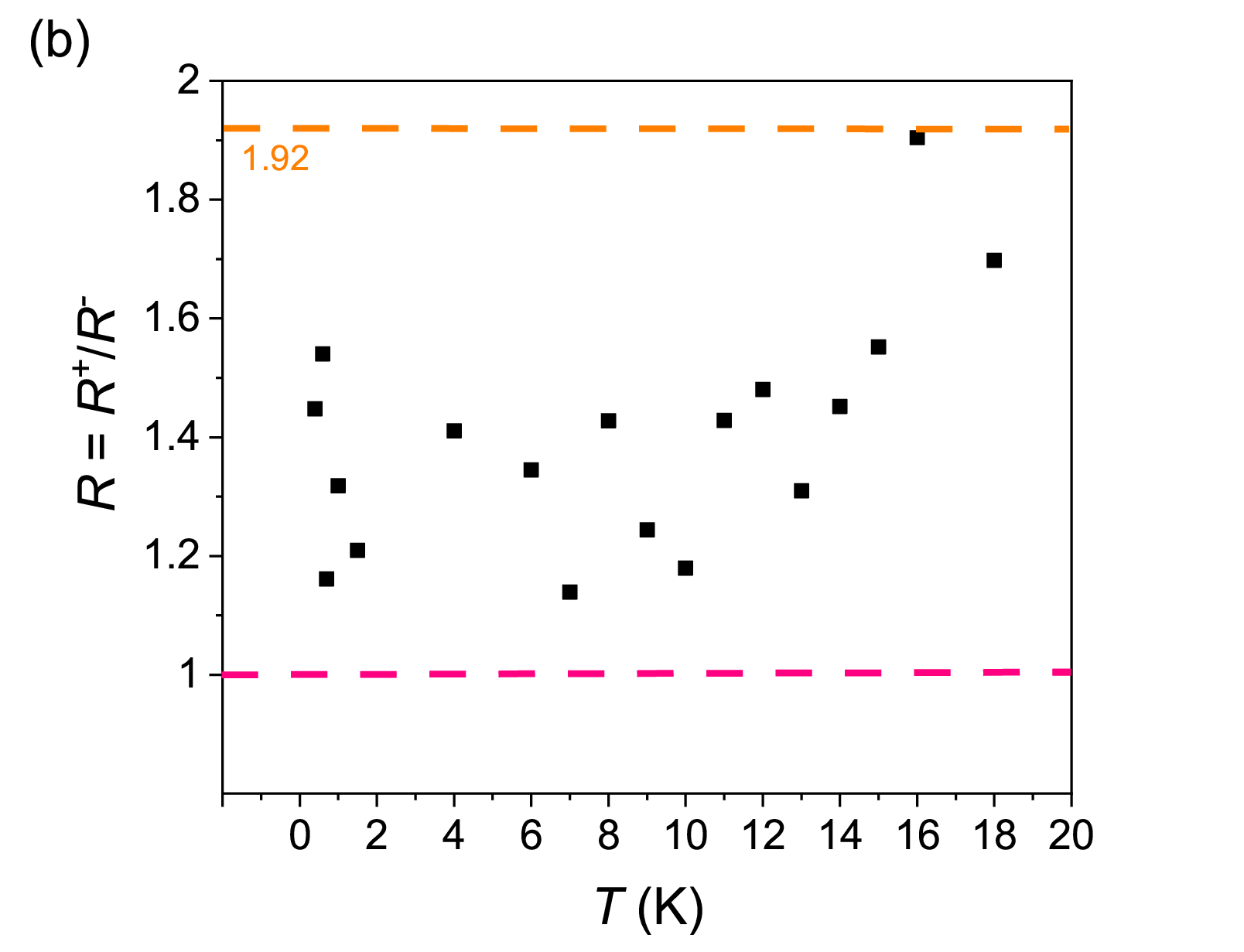}
	\caption{(a) Examination of the second derivative (red) of the pumped current (black), $d^{2}I_{p}/dV_{\rm G2}^{2}$, allows definition of the amplitude ratio $R = R^{+}/R^{-}$. (b) Measurement of $R$ with sample temperature $T$ reveals no cross-over to a thermal equilibrium regime. 
	}
	\label{fig:R_plot}
\end{figure}

Fig.~\ref{fig:R_plot}(a) plots this derivative, with the amplitudes marked as $R^{+}$ and $R^{-}$.
We note that, for a QD in equilibrium ($g = 0$), $R = R^{+}/R^{-} = 1$ \cite{Lafarge}, and for $g \rightarrow \infty$, $R \rightarrow 1.92$ for successively more coupling \cite{Yamahata3, Yamahata6}.

Evaluating the second derivative is difficult, as it is very susceptible to noise. 
However, we can see clearly we have $R^{+} > R^{-}$ and find $R \sim 1.3$.
This would imply we are in the decay cascade regime, reinforcing our first argument.
In Fig.~\ref{fig:R_plot}(b), we evaluate $R$ with sample temperature $T$ (in this case, we apply 20 point smoothing using the Savitsky-Golay filter). 
We see $R$, whilst susceptible to noise, shows no overall trend with $T$.
We would expect that, if $g \sim 1$ or lower, for $T > T_{0}$ (more accurately $gT_{0}$), $R \rightarrow 1$, thermal equilibrium would dominate.
Despite the noise, we would expect to see a downward trend in $R$.
This can be expected to be seen as the longer tail of $R^{-}$ would diminish (see Fig.~\ref{fig:R_plot}(a)), making this condition more easily identifiable \cite{Yamahata3}.
Further, we would expect this change in $R$ to occur approximately around $T = gT_{0}$ \cite{Yamahata6}, if $g$ is low, though this depends on the curvature of the QD also.
This argument allows us to assert $g \geq 5$.\\

Finally, we examine the value of $R$ itself.
We find $R \sim 1.3$, which is on the lower side of the possible range of $R$, and so is likely to be closer to $g = 5$ than a higher value, in agreement with a previous estimate \cite{Johnson3}.

To conclude this section, we take limits to the fit to eqn.~\ref{g fit}. 
The fit to eqn.~\ref{g fit} returns $T_{0} = 5.3 \pm 0.1$~K for all $g$; we take $T_{0} = 5$~K to accommodate that there is not data taken around $T = 5$~K.
Fitting with $g = 5$ returns $E_{c} = 8.03$~meV, and $E_{c} = 8.76$~meV for $g = 10$.
$E_{c} \rightarrow 9.63$~meV as $g \rightarrow \infty$. 
Based on the arguments presented in this section, we take $g = 5$ and $E_{c} = 8$~meV.

We should further note that despite the analysis of this section making some assumptions, the existence of a $T_{0}$ alone motivates eqn.~\ref{dc fit} as a most appropriate fit to infer the accuracy of the device. As discussed in Ref.~\onlinecite{Yamahata6}, using the out-of-equilibrium decay cascade may slightly overestimate the error at low $g$.

\FloatBarrier
\bibliography{reference_library_reduced}
\end{document}